\documentclass{iaus}
\usepackage{graphicx}

\title[ALFA ZOA Precursor Results] 
{The ALFA Zone of Avoidance Survey: Results from the Precursor Observations}

\author[Springob et al.]   
{C. M. Springob$^1$, P. A. Henning$^2$, B. Catinella$^3$, F. Day$^2$, R. Minchin$^3$, E. Momjian$^3$, B. Koribalski$^4$, K. L. Masters$^5$, E. Muller$^4$, C. Pantoja$^6$, M. Putman$^7$, J. L. Rosenberg$^8$, S. Schneider$^9$, L. Staveley-Smith$^{10}$}

\affiliation{$^1$Naval Research Laboratory, $^2$U. of New Mexico, $^3$Arecibo Observatory, National Astronomy and Ionosphere Center, $^4$Australia Telescope National Facility, $^5$Harvard-Smithsonian Center for Astrophysics, $^6$U. of Puerto Rico, $^7$U. of Michigan, $^8$George Mason U., $^9$U. of Massachusetts, $^{10}$U. of Western Australia}

\begin{document}

\maketitle

\begin{abstract}

The Arecibo L-band Feed Array Zone of Avoidance Survey (ALFA ZOA) will map 1350-1800 $deg^2$ at low Galactic latitude, providing HI spectra for galaxies in regions of the sky where our knowledge of local large scale structure remains incomplete, owing to obscuration from dust and high stellar confusion near the Galactic plane.  Because of these effects, a substantial fraction of the galaxies detected in the survey will have no optical or infrared counterparts.  However, near infrared follow up observations of ALFA ZOA sources found in regions of lowest obscuration could reveal whether some of these sources could be objects in which little or no star formation has taken place (``dark galaxies'').  We present here the results of ALFA ZOA precursor observations on two patches of sky totaling 140 $deg^2$ (near $l=40^{\circ}$, and $l=192^{\circ}$).  We have measured HI parameters for detections from these observations, and cross-correlated with the NASA/IPAC Extragalactic Database (NED).  A significant fraction of the objects have never been detected at any wavelength. For those galaxies that have been previously detected, a significant fraction have no previously known redshift, and no previous HI detection.

\keywords{galaxies: distances and redshifts, galaxies: ISM, radio lines: galaxies}
\end{abstract}

\firstsection 
\section{Introduction}

ALFA ZOA will include a shallow survey that covers Galactic latitudes $|b|<10^{\circ}$ and runs commensally with Galactic HI observations, and a deep survey that covers Galactic latitudes $|b|<5^{\circ}$.  While both components of the survey begin in Summer 2007, we have conducted precursor observations in 2005-2006 on two patches of sky totaling 140 $deg^2$.

The precursor observing strategy matched the observational setup of the shallow survey: Each point on the sky is covered by two passes, using the ``basketweave technique'' (where the telescope is always pointing towards the meridian, but nods back and forth in elevation) that will be utilized in the shallow survey.  The data were reduced using software originally developed for the Parkes Multibeam surveys (LiveData, Gridzilla), adapted for Arecibo.  Data were gridded using a median filter, taking advantage of the re-observations of sky pointings afforded by the basketweave technique.  Cubes were searched by eye, and HI parameters were fit in Miriad.  The parameters extracted from the spectra include sky coordinates, systemic velocity, HI flux, and HI width.  The RMS sensitivity for these observations was 5-6 mJy.

\section{Results}

72 galaxies were detected in the observed sky regions.  57 of those galaxies have a counterpart in NED within 3.5', with 32 of those having previously known redshifts in the literature.  23 of the objects have HI fluxes and widths in the literature.  The literature values come from \cite[Pantoja \etal\ (1997)]{Pantoja97}, \cite[Rosenberg \& Schneider 2000]{Rosenberg00}, \cite[Donley \etal\ 2005]{Donley05}, and \cite[Wong \etal\ 2006]{Wong06}.  Thus, even though many of our detections are previously known objects, the majority do not have previously known redshifts.  Only 10 of the 72 detections are in the inner Galaxy region (which covers 40 of the total 140 $deg^2$), where all but one of the counterparts have been detected in the radio only, with none of them included in the Two Micron All Sky Survey (2MASS, \cite[Skrutskie \etal\ 2006]{Skrutskie06}), whereas the bulk of the counterparts in the outer Galaxy region were detected by 2MASS.  We also find that the HI parameters from the literature appear to be in good agreement with our measured parameters.

\begin{figure}
 \includegraphics[height=3in,width=4in]{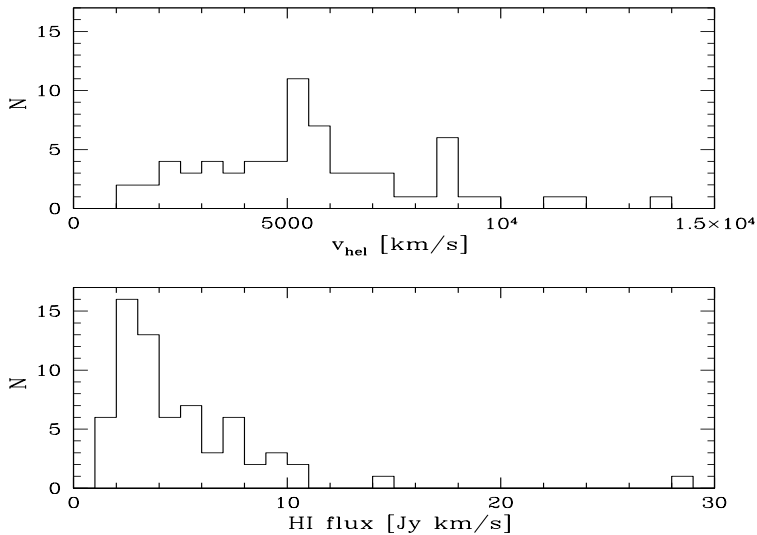}
  \caption{The distribution of redshifts ({\it top}) and HI fluxes ({\it bottom}) for all of our sources.}\label{fig:wave}
\end{figure}

\begin{acknowledgments}
This research was performed while C.M.S. held a National Research Council Research Associateship Award at the Naval Research Laboratory.  Basic research in astronomy at the Naval Research Laboratory is supported by 6.1 base funding.  P.A.H. acknowledges support from NSF grant AST-0506676.  The Arecibo Observatory is part of the National Astronomy and Ionosphere Center, which is operated by Cornell University under a cooperative agreement with the National Science Foundation.
\end{acknowledgments}

\end{document}